\documentclass[a4paper,aps,prb,reprint,showpacs]{revtex4-1}
\usepackage{CJK}
\usepackage{amsmath,amssymb,mathrsfs,amsbsy}
\allowdisplaybreaks
\usepackage[pdftex]{graphicx}
\usepackage{float} 
\usepackage{xcolor}
\usepackage{booktabs} 
\usepackage{hyperref}
\usepackage{array}

\begin{document}

\bibliographystyle{unsrt}

\title{Non-adiabatic electron-ion dynamics in anharmonic Moire structures}

\author{Shaswati Mandal}
\affiliation{Institute of Fundamental Technological Research, Polish Academy of Sciences, Adolfa Pawińskiego 5b, 02-106 Warsaw, Poland}

\author{Piotr Chudzinski}
\affiliation{Institute of Fundamental Technological Research, Polish Academy of Sciences, Adolfa Pawińskiego 5b, 02-106 Warsaw, Poland}

\date{\today}

\begin{abstract}

We study electron-lattice dynamics in twisted atomic bilayers. In these systems moiré patterns appear, which for the smallest angles, after lattice relaxation consist out of large domains of Bernal stacking separated by domain walls. These systems are a subject of intense research because they host numerous, sometimes exotic, instabilities. In this work, we investigate the role that coupling with phonons can play in this system. We identify two peculiarities. Firstly, in the presence of flat bands we expect that coupling is non-adiabatic. Secondly, because local surroundings that define the interatomic potential for each atom is different we expect that in the same system we shall have varying levels of anharmonicity. In our work we solve quantum time-evolution equations that include spatially resolved correlation between atoms and electrons motion. The correlation of motion is distinctly different for different sites. Our main result is that the strongest correlation, and thus dissipation, is present on the sites that are on the border between anharmonic and harmonic regions. These non-adiabatic effects should be taken into account when instabilities in the moire systems are explored.

\end{abstract}

 \maketitle

\section{Introduction}

After a Nobel Prize in physics for the discovery of graphene, the research on 2D materials has flourished and expanded in all possible directions. Unfortunately, bilayer graphene has turned out to be a band insulator, uninteresting from the point of view of low-dimensional collective physics. The situation has changed with a seminal paper by MacDonald \cite{Bistritzer2011}, who showed the possible existence of flat bands, heavy electrons susceptible to interactions, when a finite twist angle between the two layers is introduced. Indeed, experimental results confirmed this scenario.  Since their discovery, the moiré structures have been proven to host several orderings, and the field of research has exploded\cite{phononanharmonicityinTBM}. 

Moiré interference patterns in bilayer systems arise when two crystalline layers are rotationally stacked with respect to each other, giving rise to a long-wavelength superlattice whose structural and electronic properties depend sensitively on the twist angles\cite{Carr2020}. At small twist angles, the lattice mismatch produces a large periodic moiré pattern that folds the Dirac cones of graphene into the Brillouin zone of the moiré superlattice \cite{Andrei2020}. Previous studies have shown that twisted bilayer graphene undergoes lattice relaxation toward an energetically favorable configuration \cite{Cazeaux2023relaxation&DW}. The resulting in-plane lattice reconstruction enlarges the AB/BA stacking regions while suppressing the AA stacking regions, leading to the formation of a characteristic triangular domain structure\cite{Brown2012Twinning}\cite{Alden2013bilayergraphene}\cite{graphenebilayerwithtwist}\cite{moireintwistedbilayergraphene}\cite{Wijk_2015}\cite{Dai_TBGmoirewithtwist}.  Owing to the resulting modification of the electronic band structure, moiré materials have emerged as a versatile platform for exploring a wide range of quantum phenomena\cite{Cao2018superconductivity}\cite{Cao2018insulator}\cite{Liu2021}. In particular, twisted bilayer graphene(TBG) is predicted to exhibit flat electronic bands at a specific twist angle, commonly referred to as the 'magic angle', where the kinetic energy of electrons is strongly suppressed, and electron-electron interactions become dominant \cite{Wang_2024}\cite{Nam_2017}.

This great effort also sparked intense research on the theoretical side. Most theoretical works are focused on electron-electron interactions\cite{zang2020}; much less attention has been paid to the role of interaction with the lattice. However, it is easy to realize that flat bands represent a very unusual regime: a non-adiabatic lattice with a strong anharmonicity. The reason for it is twofold: 1) the emergence of a flat band with an electronic dispersion that is a few orders of magnitude smaller than that for graphene is one of the hallmarks of Moire bilayers\cite{Carr2020}; 2) for the AB Bernal-stacked region, just like in an ordinary graphene bilayer, the electronic liquid is a narrow-gap semiconductor while the gap goes to zero as one approaches the domain wall region\cite{Srolovitz2015}. From 1) and 2), we deduce that on the domain wall (and in its vicinity) we have localized electronic carriers with velocity comparable to the phonon velocity \cite{Non-adiabatic-2025}. Hence, the principal condition for adiabaticity is no longer valid \cite{Pisana2007}. Moreover, unlike for Bernal stacking, within the domain-wall atoms are in AA arrangement, i.e., they sit directly on top of atoms from the other layer; thus, at the maximum of the interlayer potential $V(r)$. The description ought to be given with a potential $V(r)$ of order higher than two. As a result, the domain atoms' dynamics are anharmonic \cite{Fetherolf-anharmonicphonos2023}. 

The dynamics of domain wall atoms is of a paramount importance for the physics of smallest angle bilayers. It has been shown, by adiabatic, classical methods, that large-scale reconstruction is possible in these systems\cite{Yoo2019} with Bernal stacked regions growing massively. X-ray diffraction experiments confirmed this picture. However, this means that there are very narrow regions that are far from nominal equilibrium, with large lattice reconstruction, and the question arises how stable this situation is. It is now highly desirable to asses this question by probing the quantum dynamics of atoms in these regions. Such methods are computationally expensive; thus, we decided to probe the cross section of a domain wall and its neighborhood, i.e., the spatial coordinate begins inside the domain wall and goes in an orthogonal direction towards a "normal" Bernal-stacked region.

In this paper, we solve the temporal evolution of coupled electron-lattice system and then investigate its implications for the dynamics of the electron density matrix when it is correlated with ions' motion. Such correlations are designed to capture the non-adiabatic effects while anharmonicity is incorporated with a site-dependent strength. Thus, both the non-adiabaticity and anharmonicity are captured in a way that mimics the situation inside the moire domain wall and around it. 
Our study demonstrates that lattice dynamics can significantly modify the electron density matrix through enhanced electron-phonon coupling. We further examine the momentum-dependent electron-phonon correlated motion at different sites. The spatial dependence of these correlations reveals the emergence of non-adiabatic dynamics, highlighting the role of electron-phonon correlation in the presence of anharmonicity and different coupling forces.

The outline of this paper is as follows: in Sec.\ref{sec:model} we present the model that we introduce to describe the moiré domain wall with particular focus on atomic motion at finite temperatures; in the following Sec.\ref{sec:Method} we provide a method that we use to solve the model and explain the details that we need to introduce for our specific problem. Next, in Sec.\ref{sec:results} we present a set of the most important results together with their interpretation. The paper is concluded in Sec.\ref{sec:conclusion}.

\section{Model}\label{sec:model}

\begin{figure}[h!]
        \centering
        \includegraphics[width=1.1\linewidth]{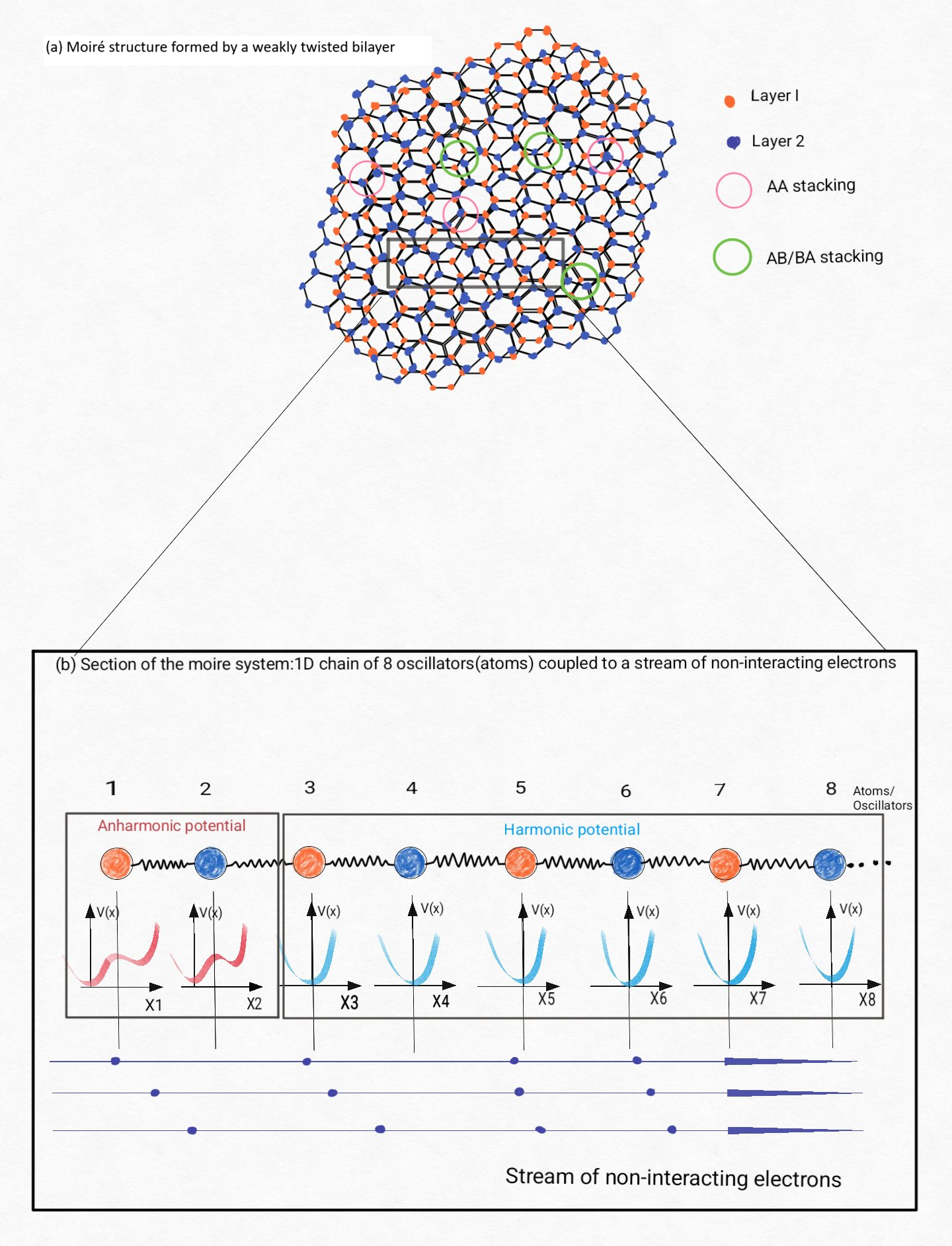}
        \caption{(a) moiré pattern formed by weakly twisted bilayers. Orange and Blue spheres denote atoms in the bottom (layer 1) and top(layer 2) layers, respectively. Regions of AA stacking where an atom of the top layer sits directly on top of the atom of the bottom layer are indicated by pink circles. Regions of AB/BA stacking where atoms in one layer sit above the void of the other layer are indicated by green circles. The rectangular section is shown in panel(b). (b) Schematic of the model used in the ECEID simulations. The highlighted region is mapped onto a 1D chain of $N_0=8$ oscillators(atoms) arranged in a chain. Atoms 1 and 2, located at the domain wall, are subject to anharmonic potentials, whereas atoms 3–8, located in the bulk of the AB/BA stacking regions, experience harmonic potentials. Each oscillator is coupled to a stream of non-interacting electrons that propagate along the chain.}
        \label{fig:placeholder}
    \end{figure}

The Hamiltonian is composed of four distinct contributions:
the electronic Hamiltonian, the bosonic(phononic) Hamiltonian, the electron-phonon interaction(perturbation term), and an inter-site coupling term describing interactions between neighboring oscillators.

\begin{equation}\label{eq:ham}
    \hat{H}_{tot}=\hat{H}_e+\sum_{\nu=1}^{N_0}\left(\frac{\hat{p}_{t\nu}^2}{2m_\nu}+\frac{1}{2}k_\nu \hat{x}_{t\nu}^2\right)-\sum_{\nu=1}^{N_0}\hat{F}_\nu \hat{x}_{t\nu}+ \hat{H}_{cl}({x}_c)
\end{equation}
Where $ \hat{H}_{cl}({x}_c)= \xi\sum_{\nu=1}^{N_0}({x^c}_{\nu}-{x^c}_{\nu{\pm} 1})^2$ is the inter-site coupling part of the oscillators with their neighboring atoms with $\xi$ is the inter-site coupling constant. \\ 

The electronic Hamiltonian $\hat{H}_e$  describes an electron initially localized at a specific site that can tunnel to either of its nearest neighbors under the attractive potential of the surrounding ions. 

 
The second term is the vibrational Hamiltonian, which accounts for the lattice dynamics, where $\hat{x}_{t\nu}$ and $\hat{p}_{t\nu}$ are the \emph{total} displacement and momentum operators of the $\nu^{th}$ atomic site, respectively, and $m_\nu$ and $k_\nu$ are the corresponding mass and harmonic force constant.
In the third term, the electron-phonon interaction is characterized by the generalized force operator $\hat{F}_\nu$. This coupling Hamiltonian mediates the non-adiabatic exchange of energy between the electron and vibrational degrees of freedom.

It will be assumed that the motion of an atom consists of a classical centroid motion $x_c$ and quantum fluctuations $x_\nu$ due to higher-energy phonon modes \cite{P.Chudzinski-2020}:
\begin{equation}
   \hat{x}_{t\nu} (t)= {x}_{c}(t)+  \hat{x}_{\nu} (t) 
\end{equation}
where the material-specific quantity for the phonon motion is due to the characteristic frequency $\omega_0$. In our approach, the specific arrangement of the two bi-layers constituting atomic lattice, that is, a peculiarity of the moiré structure, enters through the $x_{c}(t)$ trajectory. We shall assume that the low energy modes of atomic motion, e.g. acoustic phonons, are classical degree of freedom due to thermal decoherence while higher energy modes, e.g. optical phonons, are bosons that follow quantum mechanics. There is an inter-layer contribution to the inter-atomic potential $V(x)$ and it is much weaker than the intra-layer contribution. Thus, while the intra-layer term gives rise to normal modes of lattice motion (phonons), the inter-layer can be considered as a perturbation sufficiently weak to disturb mostly the low energy modes.

We introduce the microscopic model illustrated in Fig. 1., which hosts a space dependent inter-layer interaction, and investigate the role of the  $x_c(t)$ lattice dynamics in such a system. Fig.1(a) shows a weakly twisted bilayer structure in which the finite twist angle produces a moiré superlattice characterized by distinct local stacking configurations. In particular, the  AA- stacking regions correspond to local potential energy$V(x_c)$ maxima, where atoms of the two layers sit directly on top of one another, resulting in an energetically unfavorable configuration, and energy-minimum regions, where atoms of the top layer are located at the centers of the voids of the bottom layer, are the Bernal AB/BA-stacking. 
A representative section of this moiré structure is extracted to construct the reduced model shown in Figure 1(b): a one-dimensional chain of coupled oscillators, providing a reduced microscopic model of the atomic degrees of freedom coupled with a stream of non-interacting electrons. Each oscillator represents an effective atomic degree of freedom with displacement $x_j$, where $j$ labels the lattice sites. The oscillators are considered to be coupled through effective interatomic springs, while the local potential at each site accounts for the influence of its surrounding atomic environment. To distinguish the effects of local stacking variations, in this model, the atoms are divided into two groups: atoms located at the domain wall and atoms situated deep inside the bulk region.

The atoms near the domain wall experience a spatially non-uniform environment associated with the structural reconstruction and the variation of local stacking. This is represented by an anharmonic on-site potential $V(x_c)\sim {x_c}^2 +s {x_c}^4$ (for details, see Sec.\ref{sec:appendix}), which allows the restoring force to deviate from a linear dependence on displacement.
In contrast, the atoms located deep inside the bulk are assumed to experience only a harmonic potential, $V(x_c)\sim{x_c}^2$, which shows a relatively uniform and weakly perturbed lattice environment away from the domain walls. This distinction, schematically illustrated in Fig. 1(b), enables us to isolate and analyze the influence of domain wall-induced anharmonicity on the vibration of atoms of the moiré structure.

\section{Method}\label{sec:Method}
To investigate our proposed model, we employ the Effective Correlated Electron Ion Dynamics (ECEID) formalism as introduced in Ref. \cite{V.Rizzi16-ECEid} \cite{Rizzi2018}. The ECEID approach enables the study of non-adiabatic electron-ion dynamics beyond the conventional Born-Oppenheimer (BO) approximation, making it particularly suitable for describing the coupled electron-phonon dynamics in the vicinity of the domain wall.

To incorporate the effect of lattice anharmonicity, we extend the standard ECEID  equations by introducing some modifications to the ionic equation of motion. Specifically, we add finite atomic displacements from their equilibrium (centroid) position, a step that has been done in Ref.\onlinecite{P.Chudzinski-2020}, but here we ought to take different levels of anharmonicity on different sites. This modification allows us to capture the influence of nonlinear lattice vibrations on the electron-ion dynamics and to examine how anharmonicity at the domain wall atoms affects the non-adiabatic behavior of the system.

The equations are as follows:
\begin{widetext}
    
 ,

 \begin{equation}\label{eq:density}
    i\hbar\dot{\hat{\rho}}_e(t)=[\hat{H}_e,\hat\rho_e(t)]- \sum_{\nu=1}^{N_0}([\hat{F}_\nu,\hat{\mu}_\nu(t)]+x_{c\nu}(t)[\hat{F}_\nu,\hat{\rho}_e(t)]
\end{equation}

\begin{equation}\label{eq:boson}
    \hbar m\omega\dot{ \rho_{b}}(t)= Tr_e(\hat{F}_\nu \hat{\lambda}_\nu(t))- Tr_e((\hat{F}_\nu \hat{\rho}_e(t))+g_{n-local}) p^c_\nu(t)
\end{equation}

\begin{equation}\label{eq: correlation1}
    i\hbar \dot{\hat{\mu}}_\nu(t)=[\hat{H}_e,\hat{\mu}_\nu(t)]+\frac{i\hbar}{m}\hat{\lambda}_\nu(t)-\frac{\hbar}{m\omega}(\rho_{b}(t)+\frac{1}{2})[\hat{F}_\nu,\hat{\rho}_e(t)]-\sum_{\nu'=1}^{N_0}(x_{c\nu'}(t)[\hat{F}_{\nu'},\hat{\mu}_\nu(t)])
\end{equation}

\begin{multline}\label{eq: correlation2}
    i\hbar\dot{\hat{\lambda}}_\nu(t)=[\hat{H}_e,\hat{\lambda}_\nu(t)]-i\hbar\chi\hat{\mu}_\nu(t)+\frac{i\hbar}{2}\{\hat{F}_\nu,\hat{\rho}_e(t)\}-i\hbar[Tr_e(\hat{F}_\nu \hat{\rho}_e(t))+g_{n-local}]\hat{\rho}_e(t) -\sum_{\nu'=1}^{N_0}(x_{c\nu'}(t)[\hat{F}_{\nu'},\hat{\lambda}_\nu(t)])
 \end{multline}
 
\end{widetext}

where $[...,...]$ is a commutator of two quantum operators and $\omega$ is the frequency of atomic oscillations, we shall assume an optical mode with a constant frequency. 

The four coupled differential equations presented above are derived using the total Hamiltonian of the coupled electron-phonon system, Eq.\ref{eq:ham}. 

Eq.\ref{eq:density} describes the time evolution of the full density matrix $\hat{\rho}(t)$ of the coupled electron-phonon system. The dynamics of the total density matrix are governed by the Liouville equation,

\begin{equation}
    {\dot{\hat{\rho}}(t)}= \frac{1}{i\hbar}[\hat{H}_{tot},\hat{\rho}(t)]
\end{equation}
This equation provides a complete quantum-mechanical description of the coupled system of electronic and bosonic degrees of freedom.
The reduced electron density matrix is obtained by tracing over oscillator degrees of freedom.
\begin{equation}
    \hat{\rho}_e(t)= Tr_o(\hat{\rho}(t))
\end{equation}
The density matrix $\hat{\rho}_e(t)$ has all the information needed to describe the electronic system. In the absence of any other dissipation, its time evolution is governed by the Liouville equation, while the effect of electron-phonon interactions is implemented through the dynamics derived within the ECEID formalism.

Eq.\ref{eq:density} shows that the electronic density matrix evolves under the bare electronic Hamiltonian $\hat{H}_e$ and two phonon-induced terms. The first $[\hat{F}_\nu, \hat{\mu}_\nu(t)]$ involves the electron-phonon correlation operator $\hat{\mu}_\nu(t)$, accounts for correlation beyond the centroid, and the second, $x_{c\nu}(t)[\hat{F}_\nu,\hat{\rho}_e(t)]$, is a mean-field force in which electrons respond to the centroid motion of oscillator $\nu$. The anharmonicity enters in this equation through the centroid trajectory $x_{c\nu}(t)$, which, for a quartic potential, has the analytic form shown in \ref{sec:appendix}.
Initially, the electronic system is assumed to be in thermal equilibrium. Owing to the formation of a flat electronic band in the moiré superlattice, the electron occupation follows the Fermi-Dirac distribution. We then investigate how the electron density matrix evolves when the electron-phonon coupling force and the anharmonicity are varied.

Eq.\ref{eq:boson} describes the time evolution of the mean phonon(oscillator) occupation number $\rho_{b \nu}(t)$, which is a dynamical quantity characterizing the vibrational state of the lattice.
\begin{equation}
    \rho{_b}_\nu(t)= Tr(\hat{\rho{_b}}_\nu \hat{\rho}(t))
\end{equation}
Within the ECEID formalism, the mean phonon occupation evolves with the electronic density matrix through the electron-phonon coupling. In the present work, the evolution of the mean phonon occupation is used to quantify the influence of both the electron-phonon coupling and lattice anharmonicity on the vibrational dynamics of the moire system. It is assumed that the initial phonon occupation follows the Bose–Einstein distribution. To account for phonon softening in the system, phonon occupation is increased in the atoms of the domain wall, thereby modifying their initial vibrational occupation relative to the thermal equilibrium distribution.

Eq.\ref{eq: correlation1} describes the evolution of the position-dependent electron-phonon correlation operator $\hat{\mu}_\nu$, which is defined as,
\begin{equation}
    \hat{\mu}_\nu= Tr_o(\hat{x}_\nu \hat{\rho}(t))
\end{equation}
where $\hat{x}_\nu$ is the displacement operator of the $\nu$-th phonon mode, $\hat{\rho}(t)$ is the electron density matrix of the coupled system, and denotes the trace over the oscillator degrees of freedom. The operator $\hat{\mu}_\nu$  
provides a quantitative measure of the non-adiabatic coupling between the electronic and lattice systems. 

Similarly, Eq.\ref{eq: correlation2} describes the evolution of the momentum-dependent electron-phonon correlation operator $\hat{\lambda}_\nu$, which is defined as,
\begin{equation}
    \hat{\lambda}_\nu= Tr_o(\hat{p}_\nu \hat{\rho}(t))
\end{equation}
where $\hat{p}_\nu$ is the momentum operator associated with the $\nu$-th phonon mode. This operator describes the momentum correlation between the electronic and bosonic degrees of freedom. These two operators $\hat{\mu}_\nu$ and $\hat{\lambda}_\nu$ give the correlated information between the electronic system and the oscillator and play an essential role in providing a complete description of the non-adiabatic electron-phonon dynamics within the ECEID  framework. In the present work, it is assumed that the initial values of $\hat{\mu}_\nu$ and $\hat{\lambda}_\nu$ are proportional to the square of the electron-phonon coupling strength. This assumption introduces a finite degree of initial nonadiabaticity into the system, thereby making it possible to account for the initial electron-phonon correlations in the subsequent dynamic evolution.

The displacement field, $g_{n-local}$, is constructed by considering the relative displacement of each atom from its nearest neighbors.
\begin{multline}
   g_{n-local}= If[1<\nu<N_0-1, \xi(x^c_\nu(t)-x^c_{\nu+1}(t))+\\
    \xi(x^c_{\nu-1}(t)-x^c_\nu(t)),0]
\end{multline}

 Where $x^c_j$ denotes the position of the $j^{th}$ atom and $\xi$ denotes the corresponding nearest-neighbor coupling constant. The local displacement is evaluated using the positions of the neighboring atoms, $x^{c}_{j-1}$ and $x^{c}_{j+1}$, thereby accounting for the spatial correlation of the lattice deformation rather than the displacement of an isolated atom. 



 

\subsection{Hamiltonian in ECEID}

We implement the ECEID method using the following specific form of $H_{tot}$ in Eq.\ref{eq:ham}. Here, the electronic part of the Hamiltonian $H_e$ is described as
 \begin{equation}
     \hat{H}_e=\epsilon\sum_{i,j}\hat{c}_{ij}^{\dagger}\hat{c}_{ij}-t\sum_{i,j}(\hat{c}_{ij}^{\dagger}\hat{c}_{(i+1)j}+h.c) 
 \end{equation}
Here, $\epsilon$ represents the on-site energy matrix elements of the tight-binding Hamiltonian, and $t$ is the nearest-neighbor hopping constant between the atomic sites $i$ and $j$. The electronic eigenstates are obtained by diagonalizing the tight-binding Hamiltonian, and only the states in the vicinity of the Fermi level are retained in the ECEID simulation. This electronic Hamiltonian acts in the first term of Eq.\ref{eq:density}, Eq.\ref{eq: correlation1}, and Eq.\ref{eq: correlation2}.

 The electron-phonon coupling hamiltonian $\hat{H}_{e-ph}$ is represented as,
 \begin{equation}
     \hat{H}_{e-ph}=\sum_{\nu ij}{F_{\nu ij}}\hat{c_i}^{\dagger}\hat{c}_j \frac{\hat{a}_\nu^{\dagger}+\hat{a}_\nu}{\sqrt{2M_\nu \omega_\nu \hbar}}
 \end{equation}
 where $\hat{c}^\dagger(\hat{c})$ and $\hat{a}^\dagger(\hat{a})$ are the fermionic and phononic creation(annihilation) operators respectively.

 The phononic Hamiltonian $\hat{H}_{ph}$ is represented as,
 \begin{equation}
     \hat{H}_{ph}= \sum_{\nu}\hbar \omega_\nu\left(\hat{a}_\nu^{\dagger} \hat{a}_\nu+\frac{1}{2}\right)
 \end{equation}

where, in the notation of Eq.\ref{eq:ham}, $\omega_{\nu}=\sqrt{k/m}$ denotes the energy of optical phonon mode $\nu$ and we set the $\omega_{\nu}=0.35$( in the units $t$ used throughout this work).The phonon occupation follows the Bose-Einstein distribution.
\begin{equation}
    n_{\nu}=\frac{1}{e^{\hbar\omega_{\nu}^{\prime}/k_{B}T}-1},
\end{equation}
where $k_B$ is the Boltzmann constant, and T is the temperature.

The phonon energy $\omega_{\nu}^{\prime}$ depends on the atomic site. For the domain-wall atoms ($\nu=1,2$), we take $\omega_{\nu}^{\prime}=0.9\,\omega_{\nu}$, whereas for the bulk atoms ($\nu=3,\dots,8$) we use $\omega_{\nu}^{\prime}=1.5\,\omega_{\nu}$. To capture the phonon softening at the domain wall, we further increase the phonon occupation on the domain-wall atoms\cite{Phononsoftening}\cite{phononsoftening2}. However, because $\langle\mu\rangle\neq 0$,$\langle\lambda\rangle\neq 0$, and $x_c(t)\neq 0$, this Hamiltonian enfolds the reciprocal coupling between electrons and phonons. The nuclear motion modulates the electron density matrix Eq. \ref{eq:density}, and electronic occupations act back on the vibrational modes Eq.\ref{eq:boson}.

\subsection{Coupling Force}

The electron-phonon coupling force employed in this work is represented by a matrix consisting of two distinct contributions: a diagonal and an off-diagonal component:
\begin{equation}
    F= F_{dia}+F_{dis}
\end{equation}

The diagonal components modify the energy of the states through local electron-phonon interaction, whereas the off-diagonal components induce transitions between different electronic states by mixing them. The displacement force $ F_{dis}\sim\nabla_rV(r)$ can be obtained following standard formulas \cite{Displacementforce}:
\begin{equation}
    F_{dis} = ie\sum_\nu \sqrt{\frac{\hbar}{2M_\nu\omega_q}}([\epsilon_q].q)V_\nu(q)e^-iq.r^0_\nu
\end{equation}
where $\epsilon_q$ is the versor of polarization. In the basis of electron states this is the non-diagonal part $\sim q$. 

The diagonal part is taken from Ref.\cite{P.Chudzinski-2020} and it reads:
\begin{equation}
    F_{dia}=V_{di-di}(q)\frac{k_BT}{(c_{ph}q_0)}K_1\left(\frac{k_BT(k-k_F)}{(c_{ph}q_0)}\right)
\end{equation}
where $K_\nu$ is a modified Bessel function of the second kind, $q_0$ is the optical phonon mode momentum, and $c_{ph}$ denotes the velocity of the acoustic phonons, which is directly proportional to the nearest neighbour coupling constant $\xi$.

To understand their respective roles in the non-adiabatic dynamics, we shall calculate in Sec.\ref{sec:results}  the effect of the diagonal and off-diagonal coupling terms separately, as well as their combined effect.

\subsection{Chemical Potential for fermions}
Electronic reconstruction within the moiré superlattice gives rise to a flat band. At zero temperature, the electron occupation follows the Fermi-Dirac distribution.
\begin{equation}
    \langle n_i \rangle = \frac{1}{\exp\beta(E_i-\mu_c) +1}
\end{equation}
where $\beta=\frac{1}{k_BT}$,

with the chemical potential $\mu_c$ is equal to the Fermi energy at $T=0$, $\mu_c(T=0)=E_f$. At finite $T$, however, particle number conservation requires $\mu_c$ to be determined self-consistently.

The electron density of states is taken to be $g(E_i)=c\sqrt{E_i}$, where c is the proportionality constant.

The density of conduction electrons $n$ must stay fixed as $T$ varies.
\begin{equation}\label{Fermi-dirac integral}
    n(T)=\int_{0}^{\infty}dE g(E)\langle n(E)\rangle 
    =\int_{0}^{\infty}dE \frac{g(E)}{exp[\beta(E-\mu_c)] +1}
\end{equation}
which is constant. where $\beta\equiv T^{-1}$

Consequently, the chemical potential ($\mu_c$) decreases with increasing temperature to preserve the electron density. Using the Sommerfeld expansion, the leading low-temperature correction is given by \cite {AshcroftMermin1976chemicalpotential}. 
\begin{equation}
    \mu_c(T)= E_f-\frac{\pi^2}{6}(k_B T)^2\frac{\dot{g}(E_f)}{g(E_f)}
\end{equation}
which explicitly relates the temperature dependence of the chemical potential to the energy dependence of the density of states near the Fermi level.

In our present work, instead of the low-temperature Sommerfeld expansion, we determine the chemical potential $\mu_c(T)$ by solving the integration in Eq. \ref{Fermi-dirac integral} self-consistently. The Fermi-Dirac integral is expressed in a closed form through the polylogarithm function $Li_s[X]$, and the algebraic equation resulting from Eq.\ref{eq:chem-pot-poly} is solved numerically for $\mu_c$ at different temperatures: 
\begin{equation}\label{eq:chem-pot-poly}
    \mu_c(T)= E_f+\frac{Li_2[exp({-\frac{\mu_c}{k_BT}})]}{g(E_f)}
\end{equation}
This yields an accurate $\mu_c(T)$ over the entire temperature range of our simulations of a narrow band system.

\section{Results}\label{sec:results}
The four ECEID  equations are solved for the coupled electron-phonon Hamiltonian introduced in Sec\ref{sec:model}. In this paper, we focus on the time evolution of the electron density matrix $\hat{\rho}_e(t)$  from  Eq.\ref{eq:density} and of the momentum-dependent electron-phonon correlation operator $\hat{\lambda}_\nu(t)$ from Eq.\ref{eq: correlation2} under different conditions discussed below.


Using this framework, we simulate the non-equilibrium electron-phonon dynamics of the moiré model shown in Fig.1: a one-dimensional chain of eight coupled oscillators representing the atomic degrees of freedom coupled to a stream of non-interacting electrons. The electrons interact with the lattice via an electron-phonon coupling, allowing us to study non-equilibrium energy transfer between the electronic and vibrational degrees of freedom.

 The system temperature, both for the lattice and for electrons, is fixed at $T=10K$ throughout the simulation. 
\subsection{Density Matrix of Electrons}
\textbf{Case 1}: We first consider the system in the absence of both an external coupling force and lattice anharmonicity. By solving the four ECEID equations, we obtain the time evolution of the electronic density matrix. Over sufficiently long simulation times, the electronic occupations follow the Fermi-Dirac distribution.

No net energy transfer is observed between the electrons and oscillators, indicating that the electron-phonon interaction does not induce relaxation under these conditions. Consequently, the system evolves adiabatically, and the dynamics remain consistent with the Born-Oppenheimer approximation.

\textbf{Case 2}: We next consider the effect of the coupling force in the absence of lattice anharmonicity. As discussed in Sec.\ref{sec:Method}, the coupling force consists of two distinct contributions: a diagonal component $(F_{dia})$ and an off-diagonal component $(F_{dis})$. To isolate their respective roles, each contribution is activated independently while the other is kept inactive.

When only the diagonal coupling force is included, the system exhibits behavior essentially identical to that of Case 1. The electron occupations remain close to the equilibrium Fermi-Dirac distribution Figure 2(a), and no significant energy exchange is observed between the electron and lattice subsystems.

In contrast, activating only the off-diagonal coupling force leads to a pronounced modification of the electronic dynamics. Although the electronic occupations initially follow Fermi-Dirac statistics, they gradually deviate from equilibrium as the simulation progresses. In particular, electrons are pumped from the initially occupied flat band states into other electronic states, as shown in Fig. 2(b). This redistribution of the electronic population demonstrates that the off-diagonal coupling force efficiently induces interstate electronic transitions.

These results indicate that even a weak off-diagonal coupling force is sufficient to drive energy exchange between the electrons and oscillators localized at the domain wall. The diagonal coupling force alone does not produce a comparable effect, highlighting the dominant role of the off-diagonal interaction in the non-adiabatic dynamics.

\begin{widetext}

\begin{figure}[h!]
\centering
\setlength{\tabcolsep}{2pt}
  \begin{tabular}{cc}
   
\includegraphics[width=0.45\linewidth,height=0.36\linewidth]{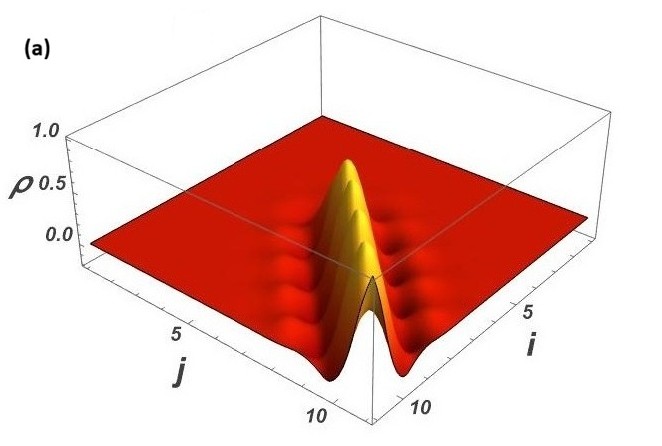}
\includegraphics[width=0.46\linewidth,height=0.37\linewidth]{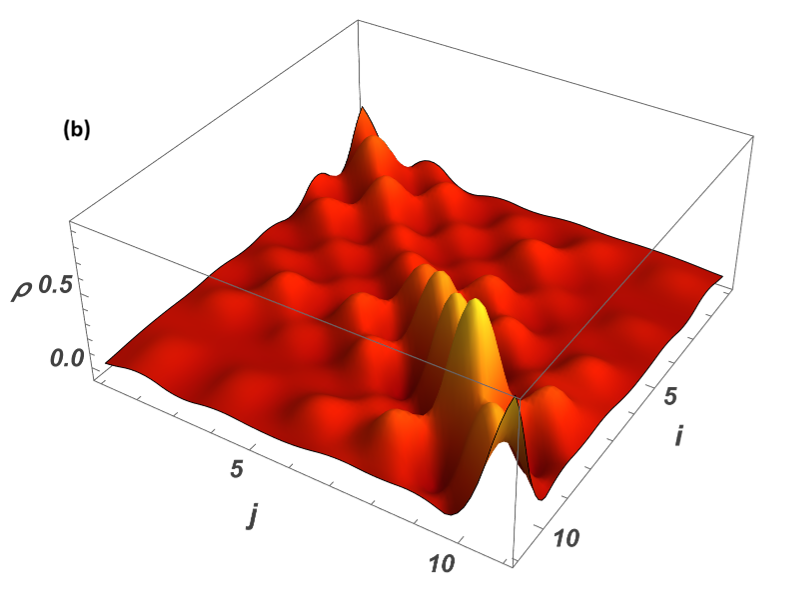}\\[4pt]
\includegraphics[width=0.45\linewidth,height=0.36\linewidth]{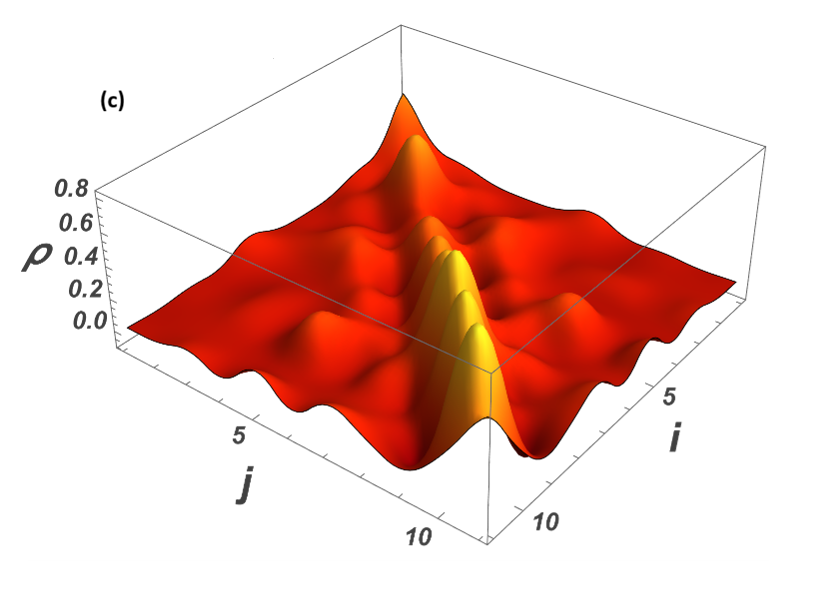}
\includegraphics[width=0.46\linewidth,height=0.36\linewidth]{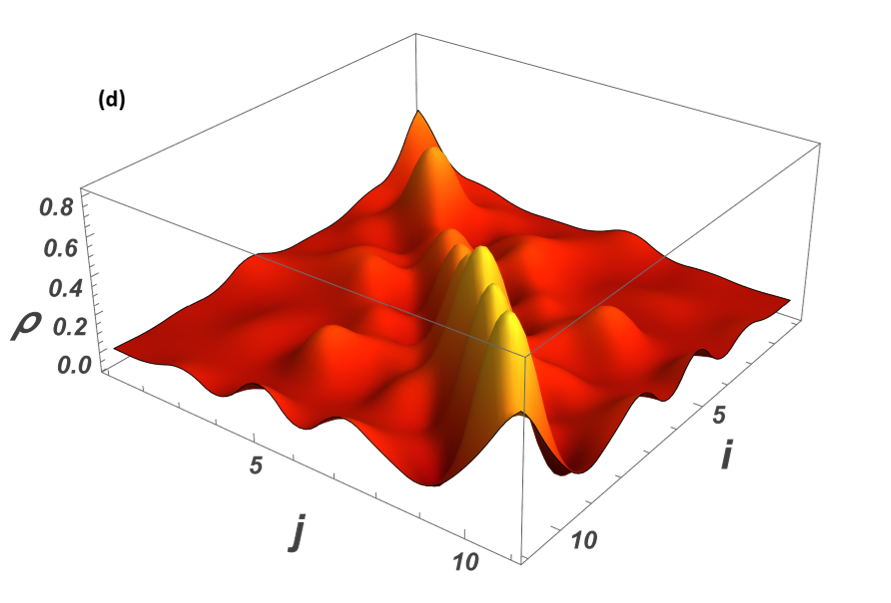}
\end{tabular}

\caption{\label{fig}{Evolution of the electronic density matrix under different contributions from the electron-phonon interaction and lattice anharmonicity. 
(a) Electronic density matrix in the presence of only the diagonal component of the electron-phonon coupling force, $ F _ {dia} $. (b) Electronic density matrix in the presence of only the off-diagonal component of the electron-phonon coupling force,$F_{dis}$. (c)Electronic density matrix when both the diagonal and off-diagonal coupling forces are included. (d) Electronic density matrix in the presence of both lattice anharmonicity and the coupling force.}}
\end{figure}
\end{widetext}

Now, when both the diagonal and off-diagonal coupling forces are activated simultaneously, the electronic dynamics are further modified, as shown in Figure 2(c). Although the diagonal coupling force alone does not influence the electron density matrix, its effect becomes significant in the presence of the off-diagonal coupling force. The interplay between the two coupling forces enhances the electronic transition, resulting in a larger population transfer from the initially occupied states to the higher-energy states compared with the case where only the off-diagonal coupling force is present. As the simulation progresses, the population of electrons in the higher-energy diagonal states continuously increases, and the characteristic Fermi-Dirac distribution gradually disappears.

This behavior indicates that the off-diagonal coupling force opens the transition channels, while the diagonal coupling further facilitates the redistribution of the electronic population.

In the absence of electron-phonon coupling, the initial nonadiabaticity is zero, and anharmonicity alone has no effect: the electronic occupations follow the normal Fermi-Dirac distribution, as in Case 1. Anharmonicity by itself therefore induces no nonadiabatic effects.


\textbf{Case 3}: In the final case, both the electron-phonon coupling and the lattice anharmonicity are simultaneously activated. Under these conditions, the electron density matrix deviates clearly from the Fermi-Dirac distribution, with increased populations in both the higher-energy diagonal states and the off-diagonal states, as shown in Figure 2(d). Although the off-diagonal elements become populated, their occupation remains significantly smaller than that of the diagonal elements, and the characteristic Fermi-Dirac distribution is no longer preserved.

A direct comparison of Figs. 2(c) and 2(d) shows that the overall distributions are qualitatively similar, with no obvious differences. To resolve the effect of the lattice anharmonicity, we calculate the difference between the two density matrices, as presented in Fig.3. The difference plot reveals a very small change of the order of $10^-3$, which is not apparent in the individual plots. In particular, the addition of lattice anharmonicity produces a slight enhancement of the off-diagonal electronic population.

\begin{figure}
\centering
\includegraphics[width=0.85\linewidth]{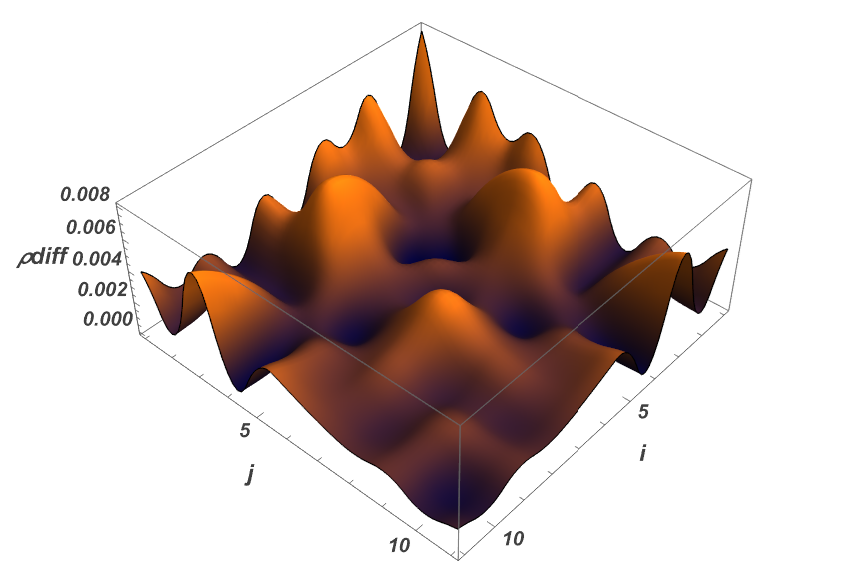}
\caption{\label{fig}{ Direct comparison between plot 2(c) and 2(d), illustrating the effect of lattice anharmonicity on the electron-phonon-induced evolution of the electronic density matrix.}}
\end{figure}

\subsection{Electron-Phonon Correlated Operator}

Eq.\ref{eq: correlation1} and Eq.\ref{eq: correlation2} describe the correlated electron-phonon operators. In particular, Eq.\ref{eq: correlation2} defines the momentum-dependent electron-phonon correlated operator, which measures the local non-adiabaticity developing in the domain-wall region. To investigate its temporal evolution, we consider two representative cases.

In both cases, the anharmonicity parameter $(s)$ is fixed at its optimal value (see \ref{sec:appendix}). In the first case, the electron-phonon coupling strength is set to zero. This configuration serves as a reference state in which the non-adiabatic contribution arising from the electron-phonon interaction is absent, and hence no coupling-induced anharmonicity is present. In the second case, a small but finite electron-phonon coupling force is introduced, thereby generating a small initial non-adiabatic contribution to the system. The subsequent time evolution is monitored to determine how the momentum- dependent electron-phonon correlated operator develops and whether this contribution is amplified during the simulation. In addition, the dynamics of individual lattice oscillators are analyzed to characterize the spatial and temporal evolution of the local electron-phonon correlation.

Particular attention is given to several representative lattice sites: the oscillators located at the domain wall, $\nu=1$, $\nu=2$, and the oscillators at the third, $\nu=3$, and fifth atomic sites, $\nu=5$, away from the wall. This analysis allows us to determine whether the non-adiabatic effects remain localized in the vicinity of the domain wall or propagate into the bulk through the coupled lattice dynamics.

For the first case, the electron-phonon coupling strength is set to zero; consequently, the initial non-adiabaticity of the system is directly proportional to the square of the electron-phonon coupling strength  Sec.\ref{sec:Method}. The momentum-dependent electron-phonon correlation operator, which remains unchanged throughout the time evolution. In the absence of electron-phonon coupling, lattice vibrations do not induce any local electronic non-adiabaticity, including in the domain wall regions. This case therefore serves as the reference state against which all subsequent results obtained for finite electron-phonon coupling are compared. 

\begin{widetext}

\begin{figure}[h!]
    \centering
 \setlength{\tabcolsep}{2pt}
  \begin{tabular}{cc}
     \includegraphics[width=0.49\linewidth,height=0.36\linewidth]{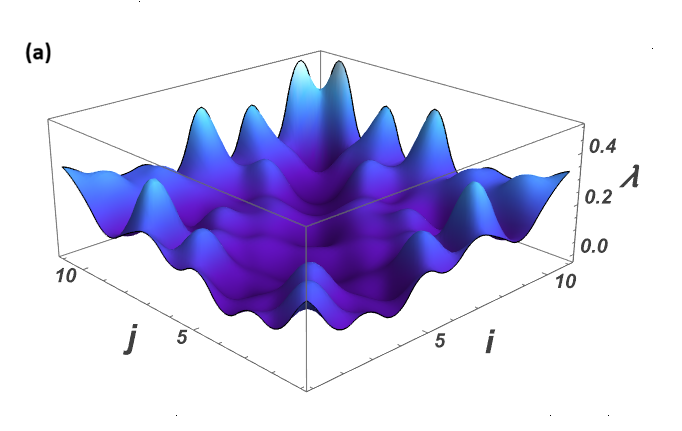}
    \includegraphics[width=0.45\linewidth,height=0.36\linewidth]{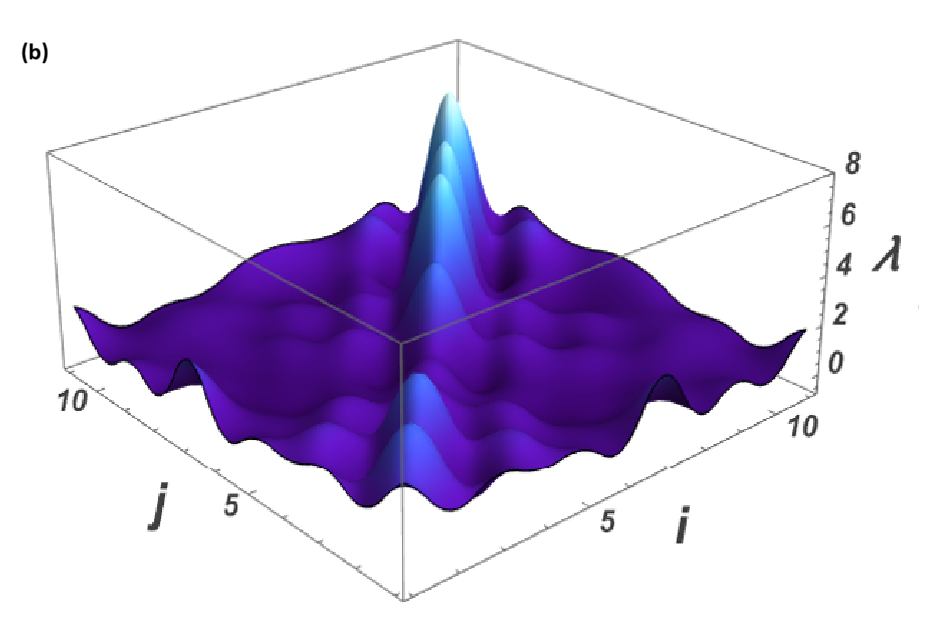}\\[4pt]
     \includegraphics[width=0.49\linewidth,height=0.36\linewidth]{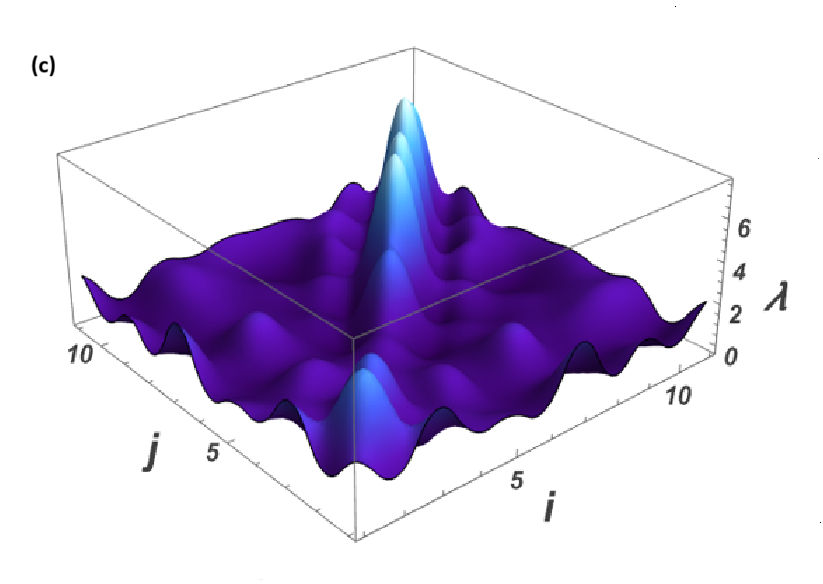}
     \includegraphics[width=0.45\linewidth,height=0.36\linewidth]{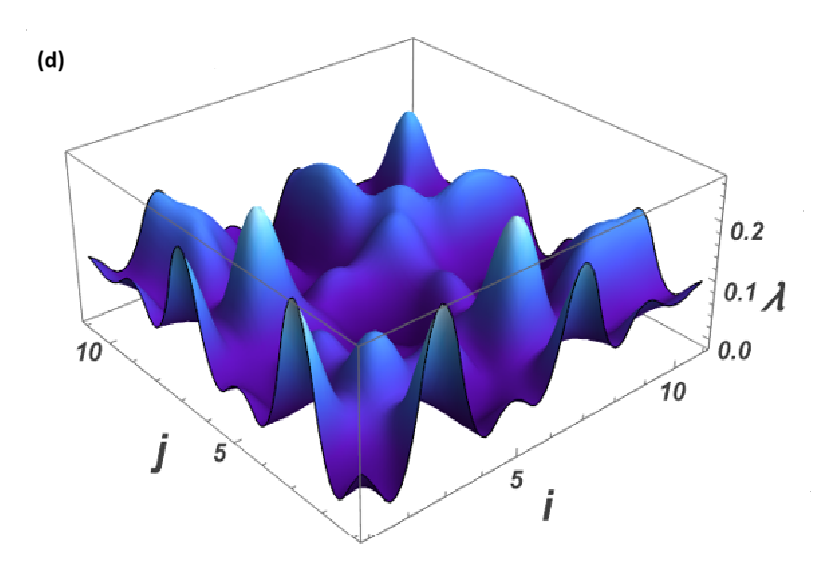}
     \end{tabular}
    \caption{Momentum-dependent electron-phonon correlation operator $\lambda_{ij}$ at a representative time $t=20$ evaluated for different oscillator sites under finite electron-phonon coupling Force $F= 0.1$ and lattice anharmonicity $s=0.81$. Panels (a)–(d) correspond to oscillator sites 1, 2, 3, and 5, respectively.
    }
    \label{fig:placeholder}
\end{figure}

\end{widetext}

For the second case, a finite electron-phonon coupling strength is introduced, resulting in a small but non-zero initial nonadiabatic contribution. Throughout this analysis, the intersite coupling constant is fixed at $\xi=0.3$. To investigate the evolution of the momentum-dependent correlated operator $(\lambda)$ with increasing distance from the domain wall towards the bulk, we consider several representative oscillator sites $ \nu=1,2,3, 5$. 
This choice of sites allows us to characterize the spatial extent of the non-adiabatic electron-phonon correlation and to determine whether its magnitude decreases, persists, or becomes enhanced as the distance from the domain wall increases.

Fig. $4(a)$ shows the temporal evolution of the momentum-dependent correlation operator $(\lambda)$ for the oscillator located at the domain wall  $(\nu=1)$ at weak but finite electron-phonon coupling. The correlation matrix exhibits a pronounced time-dependent redistribution of weight between its diagonal and off-diagonal components. Initially, the correlations are dominated by the diagonal structure, while with time an increasing off-diagonal contribution develops, reaching a pronounced maximum around $t=20$, as illustrated in Fig. $4(a)$. This redistribution signals the development of correlations between different electronic states mediated by the lattice degrees of freedom and reflects the coherent, non-adiabatic mixing of electronic and phononic dynamics. At later times, the correlation weight returns predominantly toward the diagonal, with residual off-diagonal contributions concentrated near the Fermi-level states. The resulting oscillatory transfer between diagonal and off-diagonal components demonstrates the sustained dynamical coupling between the electronic and lattice degrees of freedom and provides a direct signature of the non-adiabatic electron-phonon response.

We next consider the oscillators at and just outside the edge of the domain wall region ( $\nu=2$ and $\nu=3$); the correlation operator exhibits a qualitatively similar oscillatory evolution, with a systematic reduction in amplitude away from the domain wall. As shown in Fig. $4(b)$ and $4(c)$, the correlation weight is progressively redistributed between diagonal components concentrated near the Fermi level states and off-diagonal components associated with electronic state mixing. The diagonal contribution becomes most pronounced at $t=20$, while the subsequent enhancement of the off-diagonal weight around signals increased coherent mixing between electronic states. Although the temporal evolution is similar for both sites, the diagonal correlation amplitude is larger at site $2$ than at site $3$, indicating a gradual spatial reduction of the electron-phonon coupling away from the domain wall; just like for $\nu=1$ the subsequent periodic redistribution between diagonal and off-diagonal components reflects the dynamical exchange of momentum between state-preserving and state-mixing correlations and provides an operator-level signature of the non-adiabatic coupling between the electronic and lattice degrees of freedom.

For the fifth oscillator($\nu=5$), located farther outside the domain wall region, the correlation dynamics exhibit a qualitatively different behavior, with a substantially enhanced off-diagonal contribution compared with the oscillators near the domain wall. As shown in Fig. $4(d)$, the correlation matrix is characterized by a reduced diagonal weight and stronger state mixing in electron-phonon correlations, indicating that the coupling is increasingly distributed among electronic states away from the Fermi level. In particular, at $t=20$, the enhanced off-diagonal weight demonstrates stronger correlations between the phonon and electronic states beyond the Fermi-level sector. At later times, these off-diagonal correlations become increasingly oscillatory, indicating enhanced dynamical exchange of energy between the lattice and electronic degrees of freedom. The suppression of the diagonal component together with the growth of the off-diagonal contribution suggests that the non-adiabatic response, while weaker in overall amplitude away from the domain wall, acquires an increasingly coherent state-mixing character. This behavior indicates that the dynamical electron-phonon correlations extend into the bulk and provide an operator-level signature of the propagation of non-adiabatic electron-lattice dynamics beyond the domain wall region.

 The dynamics can be understood as an interplay between two processes: the kinetic energy associated with atomic motion and its transfer to electronic degrees of freedom through the off-diagonal components of the correlation matrix. The growth and subsequent redistribution of these off-diagonal contributions thus provide a direct signature of the time-dependent momentum exchange between lattice motion and electronic excitations.
\begin{figure}
   \includegraphics[width=0.85\linewidth]{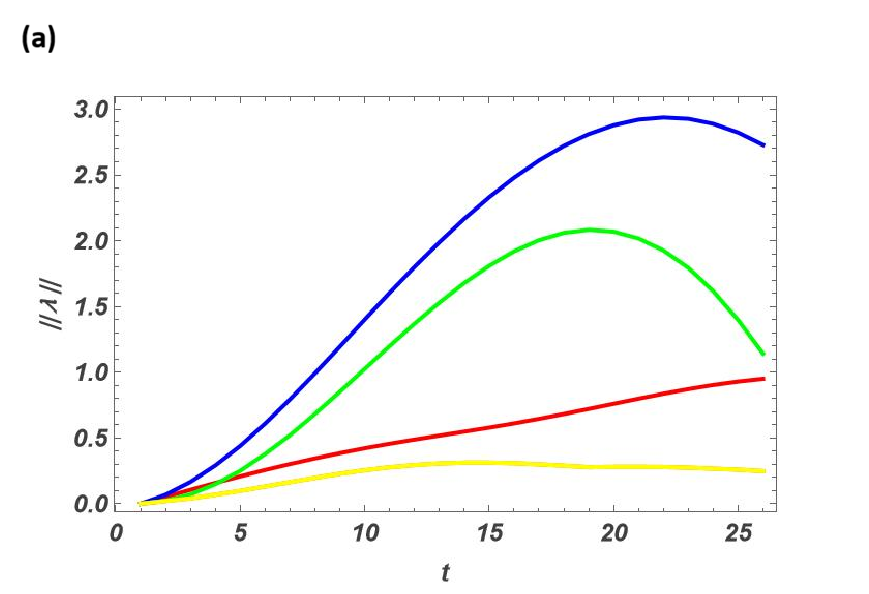}
   \includegraphics[width=0.85\linewidth]{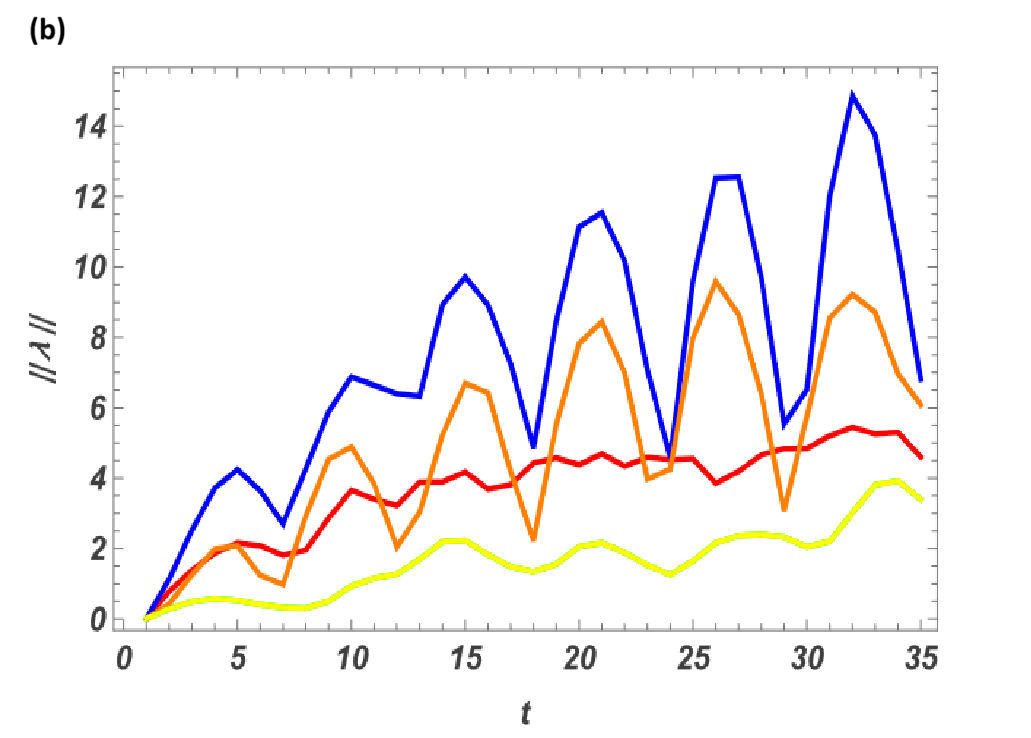}
    \caption{Time evolution of the norm of the momentum-dependent electron-phonon correlation operator at oscillator sites 1,2,3 and 5. (a) The red, blue, green, and yellow curves correspond to sites $N_0=1,2,3$ and $5$, respectively, at the electron-phonon coupling force $F=0.1$. (b) The red, blue, orange, and yellow curves correspond to sites $N_0=1,2,3$ and $5$, respectively, at the electron-phonon coupling force $F=0.3$.  }
    \label{fig:placeholder}
\end{figure}
To quantify the spatial and temporal dependence of $\hat{\lambda}$, we choose a norm $||\lambda||$ of a matrix. This norm of the momentum-dependent electron-phonon correlations operator exhibits a pronounced spatial dependence, as shown in Figure $5(a)$, at weak coupling force $F_{ep}=0.1$. The largest correlation amplitude occurs at the second oscillator site, exceeding that at the first, third, and fifth sites. This enhancement reflects the spatially inhomogeneous lattice dynamics generated by the anharmonicity, which constrains the motion of the domain wall atoms and modifies the local vibrational response. In particular, while the second oscillator is directly influenced by the anharmonic region, the neighboring third oscillator experiences the resulting dynamical distortion without being directly confined by the anharmonicity, allowing for comparatively larger local lattice fluctuations. These enhanced fluctuations increase the coupling between the phononic and electronic degrees of freedom, leading to stronger momentum-dependent electron-phonon correlations. Thus, the spatial profile of the correlation norm reflects the interplay between anharmonic lattice confinement and the resulting non-uniform electron-phonon response across the domain wall region.

We also examine the norm of the momentum-dependent electron-phonon correlation operator for the stronger coupling force, $F_{e-ph}=0.3$, and find pronounced spatial dependence and oscillatory behavior, as shown in Fig. $5(b)$. Compared with the weaker coupling force, $F_{e-ph}=0.1$ (Fig. $5(a)$), the $||\lambda||(t)$ increases substantially in amplitude and develops stronger temporal oscillations across the oscillator sites. The correlation-norm distribution also changes, indicating that stronger electron-phonon coupling alters the spatial extent and strength of the correlation response. In addition, the enhanced oscillatory behavior, including pronounced beating structure, reflects the stronger dynamical coupling between the electronic and lattice degrees of freedom. To analyze the peaks in Fig. $5$, we examine the correlation matrix, particularly for the second and third sites. For these sites, the population of the correlation matrix oscillates between the diagonal and off-diagonal elements. The peaks in the $||\lambda||$ plots correspond to maximum population along the diagonal, whereas the troughs correspond to population in the off-diagonal elements of the electron-phonon correlation matrix. This behavior is clearly shown in Fig. $6$. 

\begin{widetext}

\begin{figure}[h!]

    \centering
   \setlength{\tabcolsep}{2pt}
  \begin{tabular}{cc}
   \includegraphics[width=0.42\linewidth, height=0.36\linewidth]{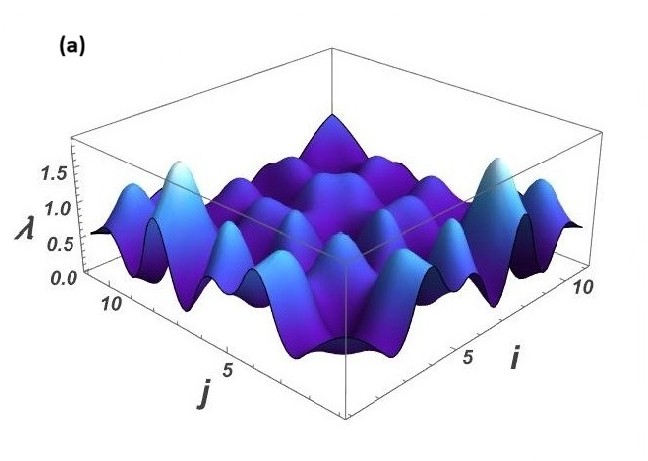}
    \includegraphics[width=0.44\linewidth,height=0.36\linewidth]{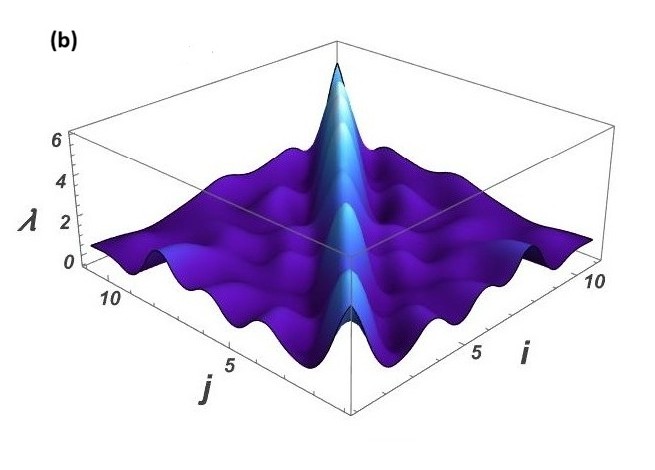}\\[4pt]
     \includegraphics[width=0.42\linewidth,height=0.36\linewidth]{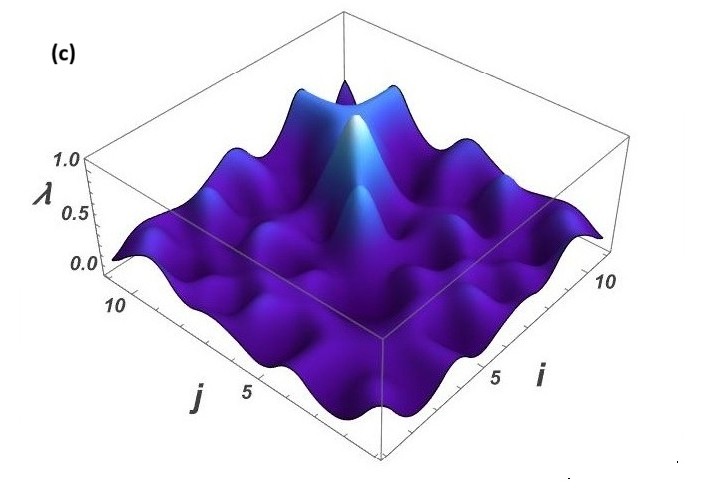}
     \includegraphics[width=0.42\linewidth,height=0.36\linewidth]{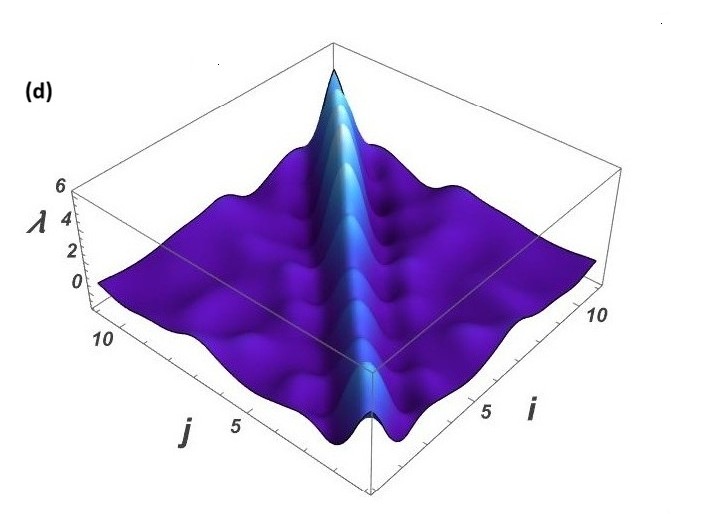}
     \end{tabular}
    \caption{Population of the momentum-dependent electron–phonon correlation matrix at two lattice sites where $||\lambda||$ is the largest and evolution times at minimum and maximum of oscillation. Panels (a) and (b) show the correlation matrix at the second lattice site for $t=18$ and $t=20$, respectively, while panels (c) and (d) present the corresponding correlation matrices at the third lattice site for the same time points. }
    \label{fig:placeholder}
\end{figure}

\end{widetext}  

This allows to put forward the following interpretation of the $||\lambda||$ oscillatory behavior. The maxima of co-motion appear when electrons with a given momentum $k$ are moving together with ions with momentum $q$, a diagonal $\hat{\lambda}$. However, when the amplitude of momentum exchange is large, this coherent state cannot persist too long. The system turns to the situation where it is the change of momenta $\Delta k = k_1 -k_2$ is co-moving with ion $q$, the off-diagonal matrix. Thus, the system is oscillating between collimated and collimating regimes. We also see that for the anharmonic site $\nu=2$, much higher transitions are active, a manifestation of phonon-phonon interactions.

\section{Conclusion}\label{sec:conclusion}

In this work, we studied the coupled dynamics of an electronic liquid and atomic lattice. We vary the anharmonicity parameter and coupling force in an effort to mimic the situation in a moiré bilayer system. In particular, we focus on non-adiabatic effects when the motion of electrons and ions becomes correlated -- a situation particularly relevant in flat bands. The most important finding of our simulations is that the non-adiabatic effects are localized and, contrary to naive intuition, they are not localized in the core of the anharmonic zone, but on the edge of the anharmonic region where it is in contact with harmonic sites, i.e. on the boundary between AA and Bernal-stacked regions. We interpret this finding in the following way. In the core of the anharmonic region, phonons suffer from a reduced lifetime, while exactly on the edge we have the best of both regions: increased co-motion thanks to a broad maximum and, at the same time, the possibility for phonon propagation not perturbed in the first order by phonon-phonon scattering (anharmonicity).

This has important implications for the physics of moire bilayers where the domain walls are present in between broad Bernal domains. According to our findings, one may expect an area of a weakly perturbed electronic liquid in the middle of the domain wall, together with a dissipative region, with hybridized electron-lattice modes surrounding the domain wall's core on both sides. For the carriers within the domain wall, we thus have a new type of one-dimensional confinement: the side motion is impeded by strong dissipation into lattice degrees of freedom. This phenomenon needs to be accounted for in future studies when properties of flat band carriers are evaluated.

\appendix

\section{Anharmonic oscillator  solution}\label{sec:appendix}

To describe the lattice dynamics, each atom is modeled as a point mass $m$ that moves without friction on a horizontal surface and is attached to a non-linear spring. The restoring force acting on each oscillator is given by
\begin{equation}
    F_x=-k x_c + sx_c^3
\end{equation}

where $k$  is the harmonic force constant, \(x_c\) denotes atomic displacement from its centroid position, and \(s\) is the anharmonic coefficient that quantifies the strength of the nonlinear interaction.

To account for the enhanced lattice nonlinearity at the domain wall, the atoms in this region are described by an anharmonic on-site potential of the form 
\begin{equation}
U(x)=\frac{1}{2} kx_c^2 -\frac{1}{4}sx_c^4    
\end{equation}
The potential is shown in Fig. $7$

\begin{figure}[h!b]
   \includegraphics[width=0.875\linewidth]{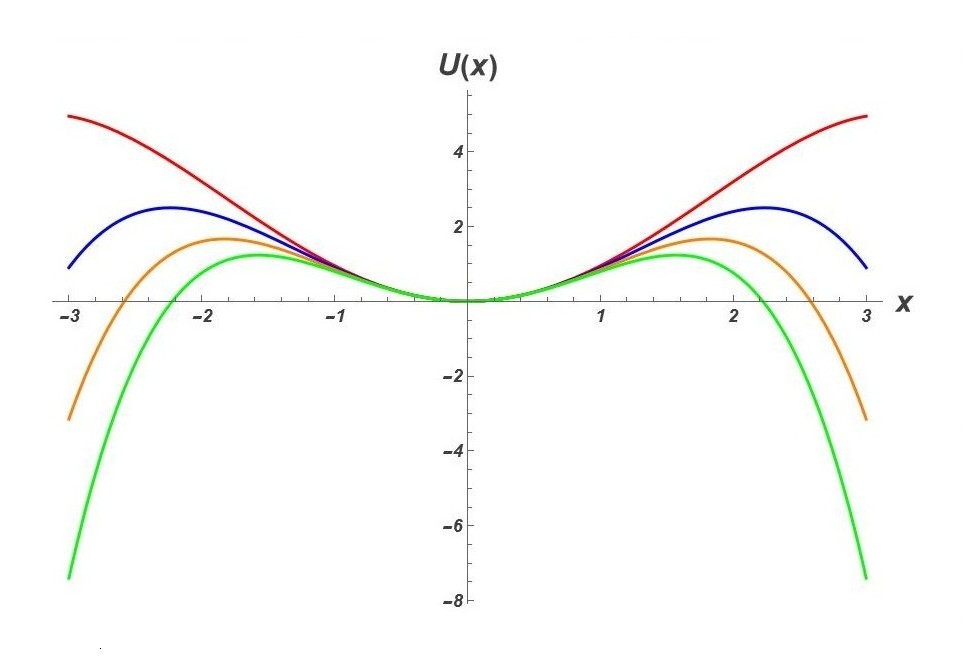}
  
    \caption{Anharmonic potential $U(x)$ at harmonic constant $k=2$, and the red, Blue, Orange, and green lines show the anharmonic coefficient $ s=0.2, 0.4, 0.6$, and $ 0.81$, respectively.}
    \label{fig:placeholder}
    \end{figure}

\begin{figure}[h!b]
 \includegraphics[width=0.875\linewidth]{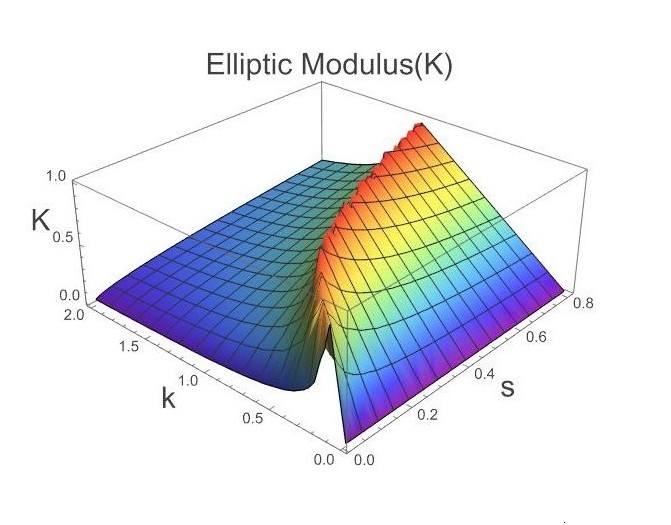}
  
    \caption{ Elliptic modulus $K$ when the energy of the oscillator $E$ is fixed at 0.42 }
    \label{fig:placeholder}
\end{figure}

The solution of equation (A1) can be expressed in terms of Jacobi Elliptic functions, which naturally arise in the dynamics of nonlinear oscillators. Depending on the nature of the motion, different elliptic functions provide physically relevant solutions. In particular, the Jacobi Elliptic functions $sn$ and $cn$ describe bounded periodic oscillations, whereas functions such as ns and dc are associated with regimes in which the restoring force becomes sufficiently weak, corresponding to the instability. In the present work, we focus on the periodic solution represented by the Jacobi Elliptic Sine function 'sn'. The corresponding elliptic modulus (K) is given by Fig.8,
\begin{equation}
    K=\frac{1}{\sqrt{2}}\left( \frac{k^2}{Es}-2-\sqrt{\left(\frac{k^2}{Es}\right)^2-4\frac{k^2}{Es}}\right)^\frac{1}{2}
\end{equation}
Here, $E$ is the energy of the acoustic phonon, taken as $E=0.42$ (in units of $ t$) in our simulation.


The displacement of the anharmonic oscillator is then expressed as,
\begin{equation}
    x(t)= a\cdot sn(b(t-t_0),K)
\end{equation}
where $t_0$ is the initial time of the evolution and the coefficients $a$ and $b$ are calculated by the equations below. 
\begin{equation}
a=\sqrt{\frac{2E}{k}(1+K^2)}
\end{equation}
\begin{equation}
    b=\sqrt{\frac{k}{m(1+K^2)}}
\end{equation}

where 'm' denotes the mass of the oscillator.\cite{Ellipticfunction}

This analytical solution provides the time-dependent displacement of the anharmonic oscillator. It forms the basis for incorporating nonlinear lattice dynamics into the modified Effective Correlated Electron Ion Dynamics (ECEID) formalism.

\paragraph*{Acknowledgments} We acknowledge interesting and inspiring discussions on exotic 2D systems with Tadeusz Burczynski. Financial support has been provided by NCN OPUS grant no. 2021/43/B/ST8/03207.

\bibliographystyle{unsrt}

\bibliography{biblioMoire}

\end{document}